\documentclass[
  preprint,          
  amsmath,amssymb,  
  aps,              
  floatfix          
]{revtex4-1}

\usepackage{graphicx}
\usepackage{amsmath}
\usepackage{amssymb}
\usepackage{bm}
\usepackage[usenames,dvipsnames]{color}
\usepackage{dcolumn}
\usepackage{hyphenat}
\usepackage{hyperref}
\usepackage{url}
\usepackage{booktabs}
\usepackage{multirow} 
\usepackage{enumitem}
\usepackage{subfigure}

\begin{document}
\title{On the tension between DESI DR2 BAO and CMB}
\author{Gen Ye}
\email[]{ye@lorentz.leidenuniv.nl}
\affiliation{Institute Lorentz, Leiden University, PO Box 9506, Leiden 2300 RA, The Netherlands}

\author{Shi-Jie Lin}
\email[]{linsj@mail.bnu.edu.cn}
\affiliation{Institute for Frontier in Astronomy and Astrophysics, Beijing Normal University, Beijing 102206, China}
\affiliation{School of Physics and Astronomy, Beĳing Normal University, Beĳing 100875, China}

\begin{abstract}
    Recent DESI DR2 baryon acoustic oscillation (BAO) measurement shows inconsistency with the cosmic microwave background (CMB) observation in $\Lambda$CDM. We used the Bayesian suspiciousness and goodness-of-fit tension metrics to quantify the significance of the tension. Both tension metrics consistently confirm a $\sim2\sigma$ tension between DESI BAO and CMB. We identified the key assumptions of $\Lambda$CDM underlying the tension between BAO and CMB. If the tension is physical, one or multiple of the assumptions might need to be broken by the potential new physics addressing the tension. In particular, the analysis shows that if dynamical dark energy is the only new physics behind the tension, phantom crossing is required to restore concordance between CMB and BAO.  
\end{abstract}
\maketitle

\section{Introduction}
The latest baryon acoustic oscillation (BAO) observation from DESI \cite{DESI:2024mwx,DESI:2025zgx,DESI:2025zpo} reports possible indication of dynamical dark energy over a cosmological constant \cite{DESI:2024aqx,DESI:2024kob,DESI:2025fii}, triggering extensive discussion in the community (see e.g. \cite{Wang:2024pui,Ye:2025ulq,Silva:2025hxw,Wang:2025ljj,Santos:2025wiv,Yang:2025mws,Li:2025cxn,Wolf:2025jed,Tyagi:2025zov,Scherer:2025esj,Liu:2025mub,Sailer:2025lxj,Jhaveri:2025neg,Ahlen:2025kat,RoyChoudhury:2025dhe,Pan:2025qwy,Akrami:2025zlb,Pan:2025psn,RoyChoudhury:2024wri}). Equally importantly, inconsistency in the standard cosmological model, the cosmological constant cold dark matter model ($\Lambda$CDM), between the CMB and DESI BAO has also been noticed with both Planck CMB \cite{DESI:2025zgx,DESI:2025zpo} and the new ACT DR6 CMB \cite{Garcia-Quintero:2025qeq}.

CMB and BAO are two of the most robust observations in cosmology up to date. After entering horizon, the primordial curvature perturbation sources perturbations in the photon-baryon plasma which propagates as sound waves, dubbed acoustic oscillations. The acoustic oscillation pattern is characterized by how far the sound waves have propagated until photon decoupling, i.e. the sound horizon. After photon decouples from electron, the acoustic oscillation pattern remains in the photon sector, which free-streams to us nearly unperturbed from recombination at $z\sim1100$, forming the main feature of the observed CMB spectra. On the other hand, the acoustic oscillation pattern frozen in the baryon sector undergoes non-linear clustering and forms the BAO signature observed in galaxy clustering at $z\lesssim2$. Therefore, CMB and BAO probe the angular scale of the sound horizon at different epochs of the Universe separated by billions of years. Together they form a baseline test of any cosmological model. 

$\Lambda$CDM, for the \textit{first} time, has shown inconsistency with this test in the recent DESI BAO observation. If confirmed, the tension between CMB and BAO would be a smoking-gun of physics beyond $\Lambda$CDM. In this paper, we used two robust and widely-adopted tension metrics in cosmology, namely \textit{Suspiciousness} \cite{Handley:2019wlz,Handley:2020hdp} and \textit{goodness-of-fit} \cite{Raveri:2018wln}, to quantify the inconsistency between BAO and CMB. 

The existence of tension has been consistently confirmed by both methods considered, with significance $2.2\sigma$ from \textit{Suspiciousness} and $2.0\sigma$ from \textit{goodness-of-fit}. Assuming the tension is physical, we discuss its possible origins by analyzing the key assumptions made in $\Lambda$CDM that are necessary to derive the tension. Specially, our analysis shows that, if dark energy is indeed the only new physics behind this tension, phantom crossing in dark energy would be inevitable.

\section{Data and method}
Following the baseline data choice of DESI \cite{DESI:2025zpo,DESI:2025zgx}, we consider the full CMB data consisting of the Camspec version of Planck PR4 high-$\ell$ TTTEEE \cite{Efstathiou:2019mdh}, Planck PR3 low-$\ell$ TTEE \cite{Planck:2019nip} and the combination of Planck and ACT DR6 CMB lensing \cite{ACT:2023dou,ACT:2023kun,Carron:2022eyg}. For BAO we use the full set of DESI DR2 data \cite{DESI:2025zpo,DESI:2025zgx} \footnote{ See also \cite{deCruzPerez:2024shj,Park:2024pew,Park:2025azv,Peng:2025nez} for some discussion using Planck PR3 and previous BAO data.}. In the rest of the paper, we will simply refer to the former as CMB and the latter as BAO.

Since we use the full CMB data, we study the standard $\Lambda$CDM model with six free parameters: cold dark matter and baryon density $\Omega_{\rm c} h^2$ and $\Omega_{\rm b} h^2$, the Hubble constant $H_0$, the primordial curvature perturbation amplitude $A_s$ and spectra tilt $n_s$ and the effective optical depth $\tau_{\rm reion}$ of reionization. Throughout we assume the Universe is spatially flat, see \cite{Chen:2025mlf} for recent discussion on the spatial curvature. We adopt the commonly used neutrino modeling of two massless species and one massive with $m_\nu=0.06$ eV, reproducing $N_{\rm eff}=3.044$ \cite{Bennett:2020zkv,Froustey:2020mcq,Akita:2020szl}. For dynamical dark energy we use the $w_0w_a$CDM model where the dark energy equation of state adopts the Chevallier-Polarski-Linder parametrization $w_{\text{DE}}(a) = w_0 + w_a(1-a)$ \cite{Chevallier:2000qy,Linder:2002et} and its perturbation approximated by the post-Friedmann framework \cite{Hu:2007pj}, which is able to restore concordance between CMB and BAO according to DESI \cite{DESI:2025zgx}. In all analyses we sample the full set of variables, $\{\Omega_{\rm c} h^2,\Omega_{\rm b} h^2, H_0, A_{\rm s}, n_{\rm s}, \tau_{\rm reion}\}$ for $\Lambda$CDM, plus $\{w_0,w_a\}$ for $w_0w_a$CDM.

To facilitate the tension estimation we sample the parameter space using the nested sampling approach implemented in \texttt{PolyChordLite} \cite{Handley:2015fda,Handley:2015vkr}, interfaced with \texttt{Cobaya} \cite{Torrado:2020dgo, 2019ascl.soft10019T}. The background and linear perturbation cosmology is computed using the standard Einstein-Boltzmann solver \texttt{CAMB} \cite{Lewis:1999bs}. $\chi^2$ minimization is done with the built-in minimizer of \texttt{Cobaya}. Suspiciousness and related quantities are computed using \texttt{anesthetic} \cite{Handley:2019mfs}. To quantify the tension between BAO and CMB, we adopt the following metric:
\begin{itemize}
    \item \textbf{Suspiciousness}        
    
    The \textit{Suspiciousness} is a prior-independent test of tension which is robust for both Gaussian and non-Gaussian posteriors \cite{Handley:2019wlz,Handley:2020hdp}. It is defined as

    \begin{equation}
        \log S = \log R - \log I,
    \end{equation}
    
    where \( R \) is the Bayes evidence ratio given by 
    
    \begin{equation} \label{BayesR}
        R = \frac{P({\rm CMB+BAO})}{P({\rm CMB})P(\rm {BAO})},
    \end{equation}
    
    where \( P({ \rm CMB}), P({\rm BAO}), P({\rm CMB+BAO})\) representing the Bayesian evidence of the model \( M, M\in\{\Lambda\text{CDM}, w_0w_a\text{CDM}\} \), for the CMB, BAO and CMB+BAO dataset respectively. If \( R \gg 1 \), it means that BAO and CMB enhance our confidence in each other by a factor of $R$ in model $M$. Conversely, if \(R \ll 1 \), we should worry of potential issues with applying $M$ to the combined dataset.  The information ratio \( I \) is defined by the Kullback-Leibler divergences \( D \) as \( \log I = D_{\rm CMB} + D_{\rm BAO} - D_{\rm CMB+BAO} \) \cite{K-L:Divergence}, accounting for the probability of two datasets matching given the prior width. The results obtained from the \textit{Suspiciousness} test allow us to evaluate the compatibility between different datasets within the context of a specific model.   
    
    Assuming the posteriors are Gaussian, the term \( d - 2\log S \), where \( d \) is the effective number of degrees of freedom constrained by both datasets, follows the \( \chi^2_d \) distribution~\cite{Handley:2019wlz}. One can then estimate the tension probability and tension significance in terms of $\sigma$ from
    
    \begin{equation}
        p = \int^{\infty}_{d - 2\log S} \chi^2_{\rm d} \, dx = \int^{\infty}_{d - 2\log S} \frac{\chi^{d/2 - 1} e^{-x/2}}{2^{d/2} \Gamma(d/2)} \, dx,\qquad \sigma(p) = \sqrt{2} \, \text{erfc}^{-1}(1 - p).
    \end{equation}
        
    \item \textbf{Goodness-of-fit loss}
    
    The goodness-of-fit \( \rm Q_{DMAP}\) quantifies the loss of explaining two datasets with the given model compared to fitting the dataset separately \cite{Raveri:2018wln}. It is defined as the difference of log likelihoods \( \mathcal{L} \) calculated at the Maximum a posteriori (MAP) parameters \( \theta_{\rm MAP} \)

    \begin{equation}
        Q_{\rm DMAP} = 2\ln \mathcal{L}_{\rm CMB}(\theta_{\rm MAP,CMB}) + 2\ln \mathcal{L}_{\rm BAO}(\theta_{\rm MAP,BAO}) - 2\ln \mathcal{L}_{\rm CMB+BAO}(\theta_{\rm MAP,CMB+BAO}).
    \end{equation}

    Assuming Gaussian posteriors, the \( \rm Q_{DMAP}\) follows a \( \chi^2_{\rm \Delta N} \) distribution with $\Delta N = N^{\rm CMB}_{\rm eff} + N^{\rm BAO}_{\rm eff} - N^{\rm CMB+BAO}_{\rm eff}$ being the difference of the respective effective number of degrees of freedom, see \cite{Raveri:2018wln} for details. The tension significance in terms of $\sigma$ can be estimated similarly to \textit{Suspiciousness}.

\end{itemize}


\section{The BAO tension in $\Lambda$CDM}

\begin{table}[h]
    \centering
    \begin{tabular}{|>{\centering\arraybackslash}p{2cm}|>{\centering\arraybackslash}p{1.8cm}>{\centering\arraybackslash}p{1.8cm}|>{\centering\arraybackslash}p{1.8cm}>{\centering\arraybackslash}p{1.8cm}>{\centering\arraybackslash}p{1.8cm}>{\centering\arraybackslash}p{1.8cm}|}
        \noalign{\hrule height 1.5pt}  
        
        \hline
        \textbf{Model}   & $\rm Q_{DMAP}$ & $\sigma_{Q_{\rm DMAP}}$ & logR & d & logS & $\sigma_{\log S}$ \\ \hline \hline
        \multirow{1}{*}{$\Lambda$CDM} 
        & $6.21$ & $2.00$ & $0.9 \pm 0.4$ & $2.4 \pm 0.8$ & $-2.7 \pm 0.2$ & $2.2 \pm 0.2$ \\ 
        \hline
        \multirow{1}{*}{$w_0w_a$CDM} 
        & 3.57 & 1.35  & $3.6 \pm 0.4$ & $4.0 \pm 0.8$ & $-0.2 \pm 0.2$ & $0.9 \pm 0.1$ \\ 
        \hline 
        \noalign{\hrule height 1.5pt}  
    \end{tabular}
    \caption{Statistical Quantities for $\Lambda$CDM and $w_0w_a$CDM.}
    \label{tab:tension metric}
\end{table}

\begin{figure}[ht!]
\centering
\includegraphics[width=0.8\textwidth]{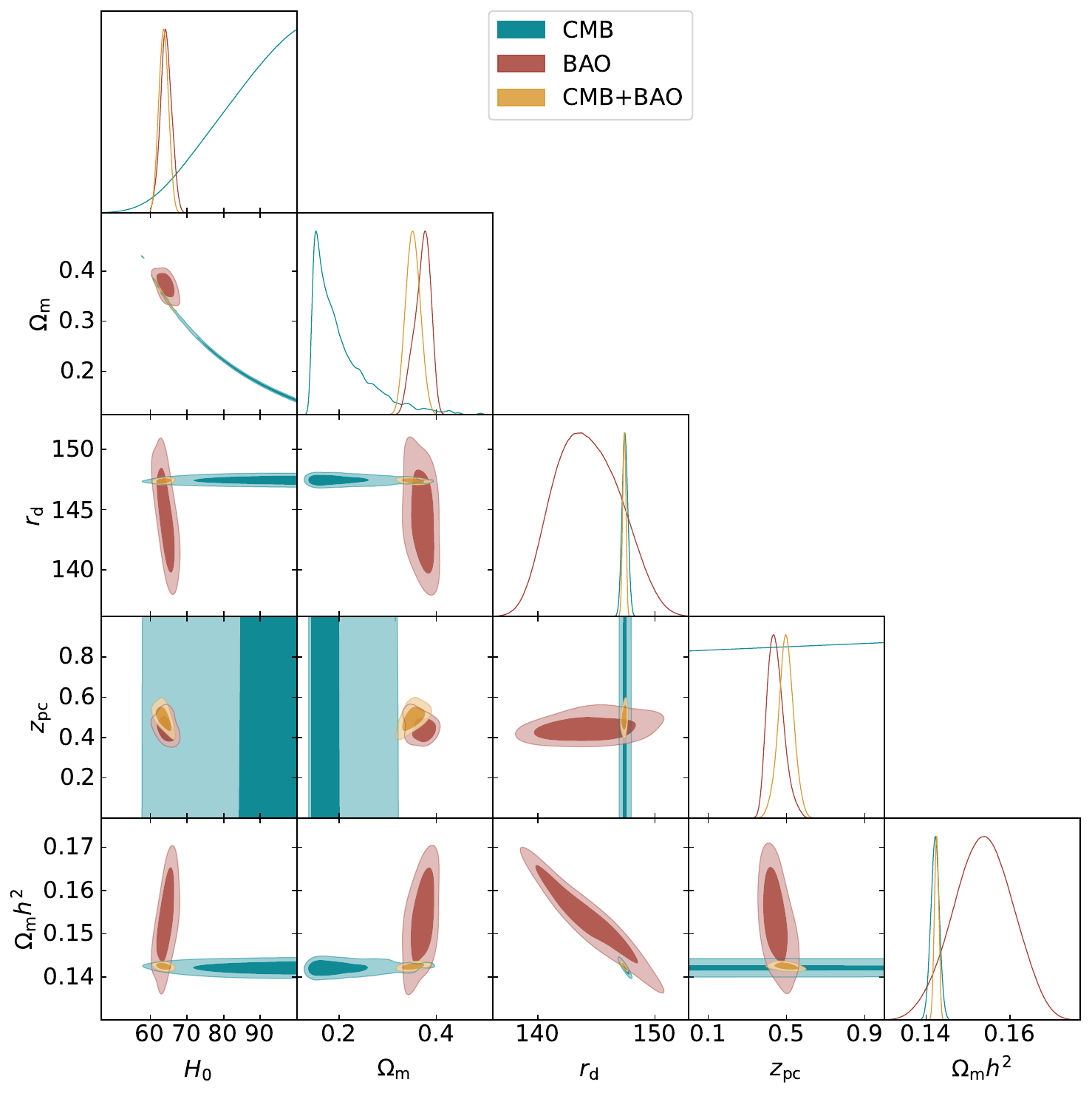}
\caption{The marginalized posterior constraint of cosmological parameters from DESI DR2 BAO (red), CMB (blue) and their combination (yellow), in $w_0w_a$CDM.
\label{fig:w0wa_contour}}
\end{figure}

Table.\ref{tab:tension metric} summaries the quantified (in)consistency between CMB and BAO in $\Lambda$CDM and $w_0w_a$CDM. The tension is confirmed by both metrics with significance $2.2\sigma$ and $2.0\sigma$ for \textit{Suspiciousness} and \textit{goodness-of-fit} respectively. This estimation is consistent with the finding of \cite{Mirpoorian:2025rfp} by directly using Monte Carlo Markov chain analysis derived posteriors. A larger than unity $\log R$ and $\log S$ consistent with 0 for $w_0w_a$CDM in Table.\ref{tab:tension metric} confirm that $w_0w_a$CDM can restore concordance between CMB and BAO. Note that the posteriors are highly non-Gaussian for the CMB only $w_0w_a$CDM, see Fig.\ref{fig:w0wa_contour} and Fig.\ref{fig:w0wa_zc}. Therefore the tension significance estimated in terms of $\sigma$, which assumes Gaussian posteriors, is not accurate for the $w_0w_a$CDM model. For $\Lambda$CDM it is still a good estimation due to the Gaussian posteriors. Alternatively, since $w_0w_a$CDM restores concordance in CMB+BAO and also includes $\Lambda$CDM as a special case $w_0=-1, w_a=0$, one can quantify the inconsistency by comparing how CMB+BAO supports $w_0w_a$CDM over $\Lambda$CDM. From the nested sampling products one can compute the Bayes factor $\ln B = \ln \mathcal{Z}_{w_0w_a\text{CDM}}-\ln \mathcal{Z}_{\Lambda\text{CDM}}=1.0$, corresponding to a $2.1\sigma$ preference of $w_0w_a$CDM over $\Lambda$CDM, consistent with the tension estimation. This is smaller than the $3.1\sigma$ reported by DESI \cite{DESI:2025zgx} estimated from the best-fit $\chi^2$ difference, see also \cite{Cortes:2025joz}.

\begin{figure}[ht!]
\centering
\includegraphics[width=0.8\textwidth]{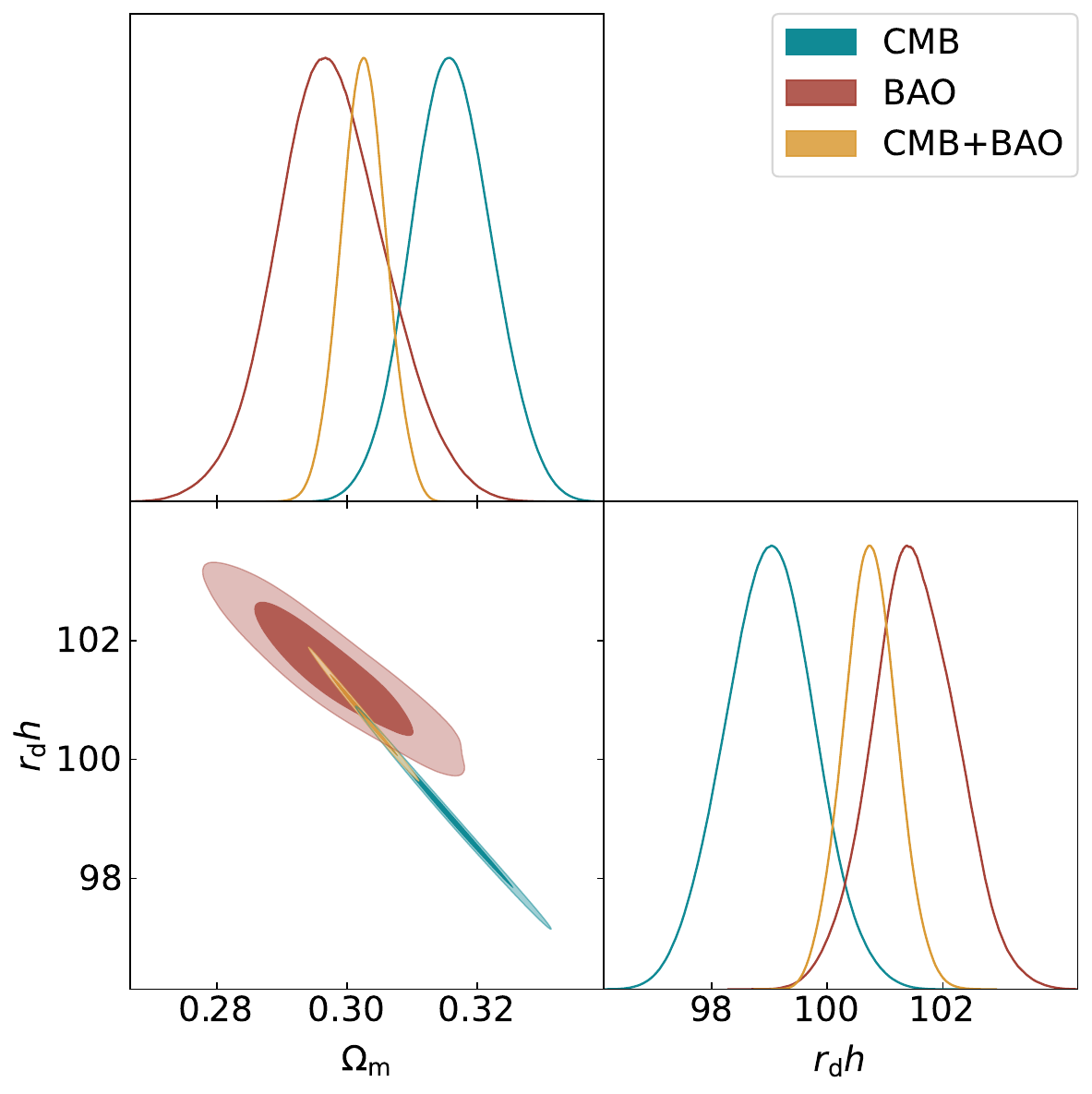}
\caption{The marginalized posterior constraint of $\Omega_{\rm m}$ and $\rm r_{d}h$ from DESI DR2 BAO (red), CMB (blue) and their combination (yellow), in $\Lambda$CDM. The dark and light shaded region mark $1\sigma \ (68\%)$ and $2\sigma \ (95\%)$ confidence level posterior constraints.
\label{fig:lcdm_omrdh}}
\end{figure}

To trace the possible origin of the tension, it is useful to investigate the assumptions of $\Lambda$CDM that facilitate the comparison of BAO and CMB. Since the observables of CMB and BAO are from two vastly different epochs (pre- and near recombination v.s. the $z<2$ late Universe) of the Universe, one naturally starts from considering what can be measured in each observation if one only makes assumption about physics in the respective epoch. BAO measures the 3D sound horizon scale in both transverse angular scale and line of sight distance difference, i.e.
\begin{equation}\label{eq:bao}
    \alpha_{\perp}(z)=\frac{r_{\rm d}}{D_{\rm A}(z)},\qquad \alpha_{\parallel}=r_{\rm d} H(z)
\end{equation}
where $r_{\rm d}$ is the sound horizon scale in the baryon sector set by CMB physics. In $\Lambda$CDM the Hubble parameter and angular diameter distance are, neglecting radiation,
\begin{equation}
    H(z) = H_0\sqrt{\Omega_{\rm m}(1+z)^3+\Omega_\Lambda},\qquad D_{\rm A}(z)=\int_0^z \frac{dz'}{H(z')}.
\end{equation}
Importantly, in $\Lambda$CDM the dark energy parameter $\Omega_\Lambda=1-\Omega_{\rm m}$ is not independent. Therefore, without making assumptions about the pre-recombination physics i.e. $r_{\rm d}$, the BAO information in \eqref{eq:bao} can be captured by the 2D posterior of $r_{\rm d}H_0 - \Omega_{\rm m}$. Fig.\ref{fig:lcdm_omrdh} visualizes the tension between BAO and CMB in the $r_{\rm d}H_0 - \Omega_{\rm m}$ plane, where the posteriors for both parameters do not match for BAO and CMB. In terms of the BAO-only constraint in Fig.\ref{fig:lcdm_omrdh}, the key assumption made in $\Lambda$CDM about the late Universe is
\begin{itemize}
    \item The background evolution is characterized by a single parameter $\Omega_{\rm m}=1-\Omega_\Lambda$.
\end{itemize}

On the other hand, to put the CMB result into Fig.\ref{fig:lcdm_omrdh}, one must make assumptions about the late Universe in addition to assuming $\Lambda$CDM during pre- and near-recombination. In fact, CMB only measures the physical matter density $\Omega_{\rm m} h^2$ in the recombination epoch by constraining the CDM density $\Omega_{\rm c} h^2$ and baryon density $\Omega_{\rm b} h^2$ at that time. The former is mainly constrained through CMB lensing and the potential envelop effect which encodes the transition from radiation dominance to matter dominance in the broad shape of the CMB temperature spectrum. The latter can be constrained by baryon drag (peak height difference between the even and odd acoustic peaks) and diffusion damping effects. By further assuming the standard recombination atomic process, the constraints on $\Omega_{\rm c} h^2$ and $\Omega_{\rm b} h^2$ determine the sound horizon
\begin{equation}\label{eq:rd}
    r_{\rm d} = \int_{z_d}^{+\infty}\frac{c_s(z')dz'}{H(z')}, \qquad c_s^2=\frac{1}{3(1+3\rho_{\rm b}/4\rho_{\rm g})},\qquad H^2(z)=\Omega_{\rm m}H_0^2(1+z)^3+\frac{\rho_{\rm g}+\rho_\nu}{3M_{\rm p}^2}
\end{equation}
where $z_{\rm d}$ is redshift of baryon decoupling. The photon density $\rho_{\rm g}$ and neutrino density $\rho_\nu$ are determined by the precise measurement of the CMB temperature \cite{Fixsen:1996nj,Fixsen:2009ug}. Therefore, \textit{CMB only directly constrains $r_{\rm d}$ and $\Omega_{\rm m}h^2$}, see their posteriors in Fig.\ref{fig:w0wa_contour}. When the post-recombination Universe is allowed to deviate from $\Lambda$CDM in $w_0w_a$CDM, Fig.\ref{fig:w0wa_contour} clearly demonstrates that CMB can still well-constrain $r_{\rm d}$ and $\Omega_{\rm m}h^2$ but looses all constraining power on $H_0$ and $\Omega_{\rm m}$. As a result, \textit{to put CMB onto Fig.\ref{fig:lcdm_omrdh} a calibration of $H_0$ is needed, which is only possible if one makes additional assumption about the post-recombination evolution}. In $\Lambda$CDM this is similar to the case of BAO in that CMB also measures the angular spacing of the acoustic peaks
\begin{equation}\label{eq:thetas}
    \theta_s=\frac{r_{\rm s}}{D_A(z_*)}
\end{equation}
where $z_*$ is the redshift of last-scattering very close to $z_{\rm d}$ thus $r_{\rm s}\simeq r_{\rm d}$ to percent level. $\Lambda$CDM needs to be assumed post-recombination so that $D_{\rm A}$ can be parameterized by only $H_0$ and $\Omega_{\rm m}\simeq\Omega_{\rm c}+\Omega_{\rm b}$. Therefore the key assumptions behind the CMB posterior in Fig.\ref{fig:lcdm_omrdh} are:
\begin{itemize}
    \item The background evolution is characterized by a single parameter $\Omega_{\rm m}=1-\Omega_\Lambda$.
    \item The correlation Eq.\eqref{eq:rd} between $\Omega_{\rm (c,b)} h^2$ and the sound horizon is unmodified.
    \item Matter and radiation density near recombination can be extrapolated to the late Universe through the standard redshift factor $a^{-3}$ and $a^{-4}$.
\end{itemize}

In general, the tension between CMB and BAO can indicate new physics breaking one or multiple of the key assumptions above. For example, there have been discussion breaking the second assumption \cite{Mirpoorian:2025rfp,Chaussidon:2025npr,Toda:2025dzd,Wang:2024dka,Pang:2025lvh,Chaussidon:2025npr}, which moves the CMB contour in Fig.\ref{fig:lcdm_omrdh} vertically to partially address the tension. Another example is to break the last assumption by decorrelating the matter density seen by CMB and BAO with negative neutrino mass \cite{DESI:2025ejh}. In the standard massive neutrino scenario, the BAO sees a slightly larger $\Omega_{\rm m}$ than the $\Omega_{\rm cb}=\Omega_{\rm c}+\Omega_{\rm b}$ from CMB due to neutrino becoming non-relativistic post-CMB, contributing to matter. Negative neutrino mass flips the effect and shifts $\Omega_{\rm m}$ to smaller values, equivalently shifting the CMB contour horizontally to the left in Fig.\ref{fig:lcdm_omrdh}, partially addressing the tension.

\section{Special case: dynamical dark energy}

Among all assumptions identified in the previous section, the first assumption, i.e. The background evolution is characterized by a single parameter $\Omega_{\rm m}=1-\Omega_\Lambda$, is special because it is the only one shared by both BAO and CMB in order to arrive at the tension in Fig.\ref{fig:lcdm_omrdh}. Breaking this assumption points to time dependence in $\Omega_\Lambda$ (effectively dynamical dark energy), or more generally \textit{unmodeled background time dependence} by $\Lambda$CDM. Interestingly, the unmodeled time dependence is also hinted by observations independent of BAO \cite{Colgain:2022nlb,Colgain:2024ksa,Risaliti:2018reu,Lusso:2020pdb,Efstathiou:2024xcq,Colgain:2024mtg,Park:2024vrw}. As shown by DESI \cite{DESI:2025fii,DESI:2025wyn} and also confirmed by the tension quantification in Table.\ref{tab:tension metric}, effectively modeling this additional time dependence as dynamical dark energy e.g. $w_0w_a$CDM, can restore concordance between CMB and BAO. It is therefore interesting to further study the possibility that only the first assumption is obviously broken.

In this case, previous discussion about CMB still holds in that it constrains $r_{\rm d}$ independently from modeling of the post-recombination physics, see Fig.\ref{fig:w0wa_contour}. Since CMB also constrains the angular acoustic scale $\theta_{\rm s}$ to 0.03\% precision \cite{Planck:2018vyg,ACT:2025fju}, Eq.\eqref{eq:thetas} implies that dynamical dark energy should be modeled in such a way that $D_{\rm A}(z_*)$ is kept approximately unchanged, see also \cite{Lewis:2024cqj}. When combined with the BAO constraint on $r_{\rm d}H_0$, this implies that dynamical dark energy must be realized in such a way that keeps the integral
\begin{equation}\label{eq:mirage}
    \int_0^{z_*}\frac{dz'}{\sqrt{\Omega_{\rm m}(1+z)^3+\Omega_\Lambda(z)}}=\frac{r_{\rm d}H_0}{\theta_s }\sim const\sim \int_0^{z_*}\frac{dz'}{\sqrt{\Omega_{\rm m}(1+z)^3+(1-\Omega_{\rm m})}}.
\end{equation}
approximately unchanged from the $\Lambda$CDM result on the RHS. On the RHS of Eq.\eqref{eq:mirage} we have the cosmological constant and the equation of state parameter $w_{\rm DE}=-1$. Therefore, that LHS equals the RHS requires that the dynamical dark energy must have an $w_{\rm DE}(z)$ that is larger than $-1$ at some period while being smaller than $-1$ in some other period such that the mean effect reproduces the RHS where $w_{\rm DE}\equiv-1$. In other words, \textit{phantom crossing is inevitable when only the first assumption is broken}. This explains why phantom crossing have been ubiquitously recovered by different analysis methods in both DESI DR1 and DR2 \cite{DESI:2024aqx,DESI:2025fii,DESI:2025wyn,Dinda:2024ktd,Jiang:2024xnu,Gao:2025ozb,Johnson:2025blf,Yang:2025kgc,Berti:2025phi,Ormondroyd:2025exu,Ormondroyd:2025iaf,Yang:2024kdo,Ye:2024ywg,Berti:2025phi,Ormondroyd:2025exu,Pang:2024qyh,You:2025uon}. It also explains why the ``mirage" parametrization \cite{Linder:2007ka} have been found by DESI to best fit the data \cite{DESI:2024kob,DESI:2025fii}. Furthermore, it also indicates that crossing the divide multiple times in an oscillatory manner might also be a possibility, see e.g. \cite{Kessler:2025kju,Specogna:2025guo}. We stress that the conclusion about inevitable phantom crossing holds when one assumes that only the first assumption is broken.

\begin{figure}[ht!]
\centering
\includegraphics[width=0.8\textwidth]{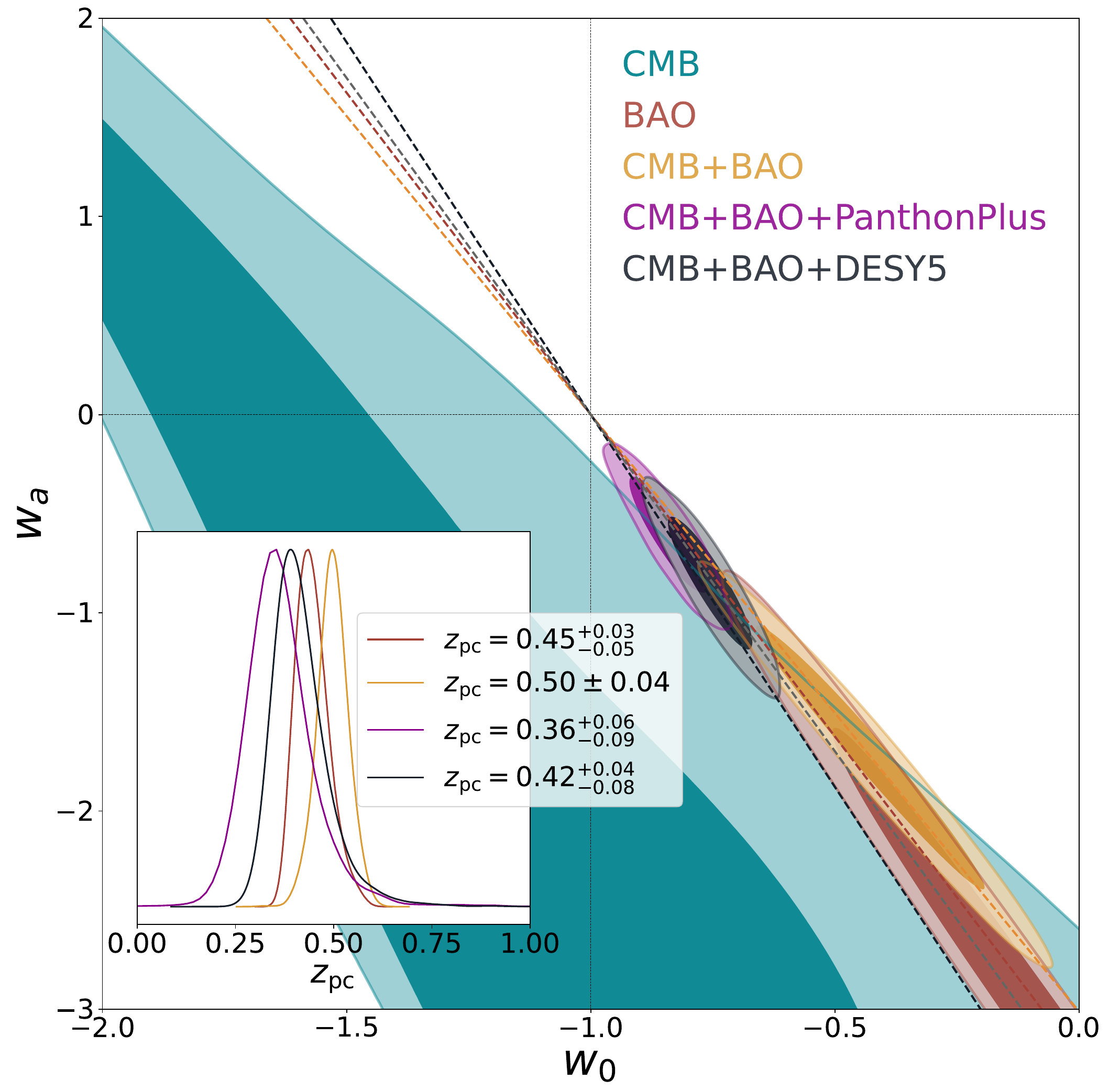}
\caption{The marginalized posterior constraints on $\omega_{0}$ and $\omega_{a}$ are shown for DESI DR2 data (red), CMB data (blue), their combination (yellow), the addition of Pantheon$+$ data (purple), and DESY5 (black), assuming a CPL dark energy model. The dashed lines indicate the mean correlation of $w_a/(w_0+1)\sim const$, which can be explained by the assumption of the phantom crossing. The inset figure displays the posterior distribution of the phantom crossing time calculated by eq. \ref{eq:crossing time}.
\label{fig:w0wa_zc}}
\end{figure}

To close this section, we note Table.\ref{tab:tension metric} indicates that data constrains only 1.5 additional degrees of freedom (d.o.f), less than the 2 new d.o.f, i.e. $w_0$ and $w_a$, introduced by $w_0w_a$CDM. This reduction in effective d.o.f corresponds to the degeneracy direction $w_a/(w_0+1)\sim const.$ in the $w_0-w_a$ plane, see Fig.\ref{fig:w0wa_zc}. Interestingly, the constrained direction also corresponds to the phantom crossing time $z_{\rm pc}$ in $w_0w_a$CDM, i.e.
\begin{equation}\label{eq:crossing time}
    z_{\rm pc} = -\frac{w_0+1}{w_a+(w_0+1)}.
\end{equation}
As depicted in both Fig.\ref{fig:w0wa_contour} and Fig.\ref{fig:w0wa_zc}, $z_{\rm pc}$ is well-constrained by the BAO data alone, yielding $z_{\rm pc}=0.45\pm0.04$. CMB contributes almost no additional constraining power to $z_{\rm pc}$ as shown in Fig.\ref{fig:w0wa_contour}. Note this correlation direction is robust even if the contour contains the $\Lambda$CDM point, as is the case in various type Ia Supernova (SN) observations \cite{Brout:2022vxf,DES:2024jxu,Rubin:2023ovl} as well as previous BAO measurements \cite{BOSS:2016wmc,eBOSS:2020yzd}. The constraint on $z_{\rm pc}$ from SN is slightly different but still statistically compatible $z_{\rm pc}$ from BAO, see Fig.\ref{fig:w0wa_zc} for the result with Pantheon+ \cite{Brout:2022vxf} and DESY5 \cite{DES:2024jxu}. The discussion about possible (in)consistency with different SN datasets is also important but out of scope of this paper, see e.g.\cite{Peng:2025nez,Ishak:2024jhs,Efstathiou:2024xcq,Huang:2025som,Ormondroyd:2025iaf,Ormondroyd:2025exu}. 

\section{Conclusion}
In this paper we quantified the tension between CMB and latest DESI DR2 BAO to be $\sim2\sigma$ with both the \textit{Suspiciousness} and \textit{goodness-of-fit} tension metrics. We identified three key assumptions of $\Lambda$CDM that this tension relies on, namely:
\begin{itemize}
    \item The background evolution is characterized by a single parameter $\Omega_{\rm m}=1-\Omega_\Lambda$.
    \item The correlation Eq.\eqref{eq:rd} between $\Omega_{\rm (c,b)} h^2$ and the sound horizon is unmodified.
    \item Matter and radiation density near recombination can be extrapolated to the late Universe through the standard redshift factor $a^{-3}$ and $a^{-4}$.
\end{itemize}
If the tension is physical, these assumptions represent the potential entry point of new physics to address the tension. Furthermore, the first assumption is special among the three in that it is the only one that is required to derive the tension from both BAO and CMB. The special case of breaking only the first assumption is introducing dynamical dark energy. For this special case we found that dark energy must cross the phantom divide in order to address the tension. In particular, we found that BAO data alone effectively constrains the time of phantom crossing $z_{\rm pc}=0.45\pm0.04$. A similar observation has been made in \cite{Scherer:2025esj} appearing on \texttt{arXiv} while we were finalizing this paper.

Phantom crossing is of profound theoretical importance in that it is usually associated with catastrophic instabilities \cite{Carroll:2003st,Cline:2003gs,Dubovsky:2005xd,Creminelli:2008wc}. It is found that to stably explain the observed phantom crossing, there is strong indication that gravity should be modified to be non-minimally coupled with matter \cite{Ye:2024ywg,Wolf:2024stt,Pan:2025psn,Wolf:2025jed,Wolf:2025jed}. Interestingly, the covariant gravity theory, \textit{Thawing Gravity}, stems from this observation naturally resolves both the BAO tension and the Hubble tension simultaneously, while keeping consistent with larger scale structure observations \cite{Ye:2024zpk}. Finally, we stress that \cite{Ye:2024zpk} demonstrates the importance of a consistent dynamical description of the dark energy/modified gravity field in both the early and late Universe, otherwise we may risk making incorrect conclusions. For example, if one only focuses on dynamics in the late Universe with either phenomenological parametrizations of $w_{\rm DE}(a)$ \cite{DESI:2025fii} or a covariant Lagrangian \cite{Ye:2024ywg,Wolf:2024stt,Wolf:2025jed}, one obtains a $H_0$ lower than $\Lambda$CDM that seemingly escalates the Hubble tension, while the full model, after consistently considering its dynamics at all times, is in fact consistent with local $H_0$ measurement \cite{Ye:2024zpk}.

\begin{acknowledgments}
The authors thanks Alessandra Silvestri and Bin Hu for discussions and proof reading of the manuscript. G. Y. acknowledges support by NWO and the Dutch Ministry of Education, Culture and Science (OCW) (grant VI.Vidi.192.069). S. L . is supported by the China Manned Space Program with grant no.CMS-CSST-2025-A04. Some of the plots were made with \texttt{GetDist} \cite{Lewis:2019xzd}. The authors acknowledge computational support from the ALICE cluster of Leiden University.
\end{acknowledgments}

\bibliography{reference}

\end{document}